\documentclass[aps,prd,onecolumn,nofootinbib,superscriptaddress,groupedaddress,preprintnumbers]{revtex4}
\usepackage{graphicx}  
\usepackage{bm}        
\usepackage{amssymb,amsmath}   
\usepackage{subfig}
\usepackage{enumerate}
\usepackage{color}
\input{colordvi.tex}

\newcommand{\nc}{\newcommand}
\nc{\hf}{\frac{1}{2}}
\nc{\bea}{\begin{eqnarray}}
\nc{\eea}{\end{eqnarray}}
\nc{\be}[1]{\begin{equation} \mbox{$\label{#1}$}}
\nc{\ee}{\vspace{0.1cm}\end{equation}}
\nc{\eq}[1]{\mbox{Eq.\ (\ref{#1})}}
\nc{\fig}[1]{\mbox{Fig.\ (\ref{#1})}}
\nc{\fb}[2]{\left(\frac{#1}{#2}\right)}

\def\GeV{{\rm \ GeV}}
\def\MeV{{\rm \ MeV}}

\def\TeV{{\rm \ TeV}}

\nc{\fnl}{f_{\rm NL}}
\nc{\gnl}{g_{\rm NL}}

\nc{\mpl}{M_{P}}

\nc{\Geff}{\Gamma_{eff}}
\nc{\Hub}{H_{\ast}}
\nc{\siga}{\sigma_{\ast}}

\def\Mpl{M_{\rm pl}}
\def\Hinf{H_{*}}

\begin{document}

\title{The minimal curvaton-higgs model}
\author{Kari Enqvist}
\email{kari.enqvist@helsinki.fi}
\affiliation{Physics Department, University of Helsinki and Helsinki Institute of Physics\\
P.O. Box 64, FI-00014, Helsinki, Finland}
\author{Rose N. Lerner}
\email{rose.lerner@desy.de}
\affiliation{Physics Department, University of Helsinki and Helsinki Institute of Physics\\
P.O. Box 64, FI-00014, Helsinki, Finland}
\affiliation{Deutsches Elektronen-Synchrotron DESY, 22607 Hamburg, Germany}
\author{Tomo Takahashi}
\email{tomot@cc.saga-u.ac.jp}
\affiliation{Department of Physics, Saga University, Saga 840-8502, Japan}

\preprint{HIP-2013-21/TH, DESY13-180}

\begin{abstract}

We present the first full study of the minimal curvaton-higgs (MCH) model, which is a minimal interpretation of the curvaton scenario with one real scalar coupled to the standard model Higgs boson. The standard model coupling allows the dynamics of the model to be determined in detail, including effects from the thermal background and from radiative corrections to the potential. The relevant mechanisms for curvaton decay are incomplete non-perturbative decay (delayed by thermal blocking), followed by decay via a dimension-5 non-renormalisable operator. To avoid spoiling the predictions of big bang nucleosynthesis, we find the ``bare'' curvaton mass to be $m_\sigma \geq 8\times 10^4\GeV$. To match observational data from Planck there is an upper limit on the curvaton-higgs coupling $g$, between $10^{-3}$ and $10^{-2}$, depending on the mass. This is due to interactions with the thermal background. We find that typically non-Gaussianities are small but that if $\fnl$ is observed in the near future then $m_\sigma \lesssim 5\times 10^9\GeV$, depending on Hubble scale during inflation. In a thermal dark matter model, the lower bound on $m_\sigma$ can increase substantially. The parameter space may also be affected once the baryogenesis mechanism is specified.

\end{abstract}

\maketitle

\section{Introduction}

In the curvaton scenario \cite{curvaton}, the primordial density perturbations are generated after a period of primordial inflation, the details of which need not be specified\footnote{
The spectral index $n_s$ is determined by both the inflaton and the curvaton potential. Although $n_s$ is now severely constrained by Planck data \cite{Ade:2013uln}, we do not calculate it because we have not specified the inflaton potential. For the discussion on the spectral index in the curvaton or a spectator field model, see \cite{Kobayashi:2013bna}.
}. Once inflation is over and the inflaton has decayed into radiation, which we assume to consist of the standard model degrees of freedom, the curvaton field begins to evolve. Eventually it starts to oscillate in its potential, and in the process the inflationary perturbations imprinted on the curvaton field grow. The final amplitude of the perturbation, as well as its statistical properties, is defined by the time of the curvaton decay. The decay products inherit the field perturbation, which is then converted into an adiabatic curvature perturbation by subsequent thermalisation. Recently, attempts have been made to determine the curvaton decay width in terms of model parameters, and also to connect the curvaton model with the standard model and other particle physics scenarios \cite{Enqvist:2012tc,Enqvist:2013qba}. Here the assumption is that the curvaton provides all of the observed perturbation; for scenarios where both the inflaton and the curvaton contribute, see \cite{Langlois:2004nn}.

In order to obtain the observed curvature perturbation amplitude of $\zeta = 4.7 \times  10^{-5}$ \cite{Ade:2013zuv}, the oscillating homogeneous curvaton condensate must be relatively long lived. Moreover, to obtain realistic estimates for the properties of the curvature perturbation, one should know both the form of the curvaton potential and the mechanism by which the curvaton decays. In addition to the amplitude of the curvature perturbation, non-Gaussianity can also give a stringent constraint on the curvaton model \cite{Lyth:2005du}. Recent Planck data has provided a constraint on a non-linearity parameter $\fnl$, which represents the amplitude of the bispectrum of the curvature perturbation. This constraint is $ -8.9 < \fnl < 14.3$ (95\% C.L.) \cite{Ade:2013ydc} for the local type, which excludes a large parameter space of the curvaton model with a quadratic potential \cite{Ade:2013ydc}. However, curvaton self-interactions may change the naive expectations for non-Gaussianity parameters considerably \cite{Enqvist:2005pg} and the extent of the exluded region in the parameter space in such a case remains to be studied.

Our aim in this paper is to consider carefully the predictions from a very specific implementation of the curvaton scenario, consisting of the standard model plus one curvaton field that is coupled to the Higgs boson. We will limit ourselves to the simplest quadratic curvaton model, but will account for the effective self-interaction arising through radiative corrections induced by the curvaton-higgs coupling. With such a specific model, we are able to calculate the decay width of the curvaton using particle physics. The dominant forms of decay are resonant production of higgses and decay to the standard model via dimension-5 operators. Using this information, we are able to make accurate predictions for the non-Gaussianity parameters in this model.

The structure of this paper is as follows. In the next section, we describe the MCH model. We give the expressions for the potential and the decay rates, which are necessary to calculate the density perturbation. Then in Section~\ref{sec:perturbation}, we summarize the formalism to calculate the curvature perturbation and its correlation functions such as the power and bi- spectra, which will be used to test the model with observational data from Planck. In Section~\ref{sec:results}, we give a prediction  for the power spectrum and non-linearity parameter $\fnl$  and discuss the constraints on the parameters in the model. The final section is devoted to our conclusions.

\section{MCH model to one loop}
\label{sec:formalism}

In order to produce the observed adiabatic curvature perturbation, the curvaton must at some point decay to radiation. Therefore it must be coupled to other fields. Assuming a gauge singlet curvaton, the only renormalizable way to couple it to the standard model is via the Higgs field. The simplest possibility is then given by the potential
\be{V_0}
V_0 = \hf m_\sigma^2 \sigma^2+ g^2 \sigma^2\Phi^\dagger \Phi~,
\ee
where $g$ is a free coupling constant, and   $\Phi$ is the complex Higgs doublet. Within a few thousand Hubble times after inflation, the higgs settles to the minimum of its potential \cite{Enqvist:2013kaa}. Thus, for the purposes of this paper, we can consider only the physical (real) higgs field, which has a normalisation of $1/\sqrt{2}$ compared to $\Phi$. In principle, we could also have assumed a trilinear coupling of the type $ \sigma \Phi^\dagger \Phi$, but this could have been done only at the cost of introducing a new mass scale. Such a term can be forbidden by imposing the global symmetry $\sigma\to -\sigma$ in the scalar sector. 
The dimension-5 operator introduced later (\eq{dim5}) does not respect the symmetry $\sigma\to - \sigma$ of the potential (\eq{V_0}). However, it is assumed that gravity breaks all global symmetries. We have also assumed the bare quartic coupling is negligible (an effective quartic coupling is generated by loop corrections). With this setup, we now can calculate the effective potential and the decay rate for the curvaton, including quantum and thermal corrections.

\subsection{Scalar potential}
Both thermal corrections and quantum correction of the Coleman-Weinberg type exist because of the curvaton-higgs coupling in (\ref{V_0}). (There are also contributions from curvaton-only terms.) Thus the simplest complete curvaton potential reads
\be{V_eff}
V_{eff}(\sigma, T) = V_0(\sigma) + \Delta V(\sigma) + V_T(\sigma, T),
\ee
where
\be{thermalcorr}
V_T=\frac {1}{24} g^2T^2\sigma^2,
\ee
is the (one-loop) thermal correction due to physical higgs loops, valid when the effective higgs mass\footnote{
In principle the dependence of the thermal mass on temperature is more complicated (see e.g.\ \cite{Drewes:2013iaa}).
} $m_H\ll T$. The one-loop Coleman-Weinberg correction is given by
\be{deltaV}
\Delta V(\sigma) =  \frac{\left(g^2\sigma^2+m_h^2\right)^2}{64\pi^2} \log\left(\frac{g^2\sigma^2+m_h^2}{\mu^2}\right),
\ee
where $m_h = 126\GeV$ is the higgs mass excluding interaction terms and $\mu$ is the renormalisation scale. For our numerical studies, we choose $\mu = m_h$. This is an arbitrary choice in principle, but different values cause the couplings to be defined at different scales. We do not consider the running of couplings here, because it is a higher-order effect than the existence of the Coleman-Weinberg potential.

Given the curvaton potential (\ref{V_eff}), one could define an effective decay rate of the curvaton, $\Gamma_{eff}$, which is a dynamical quantity that depends parametrically on the initial curvaton field value $\sigma_*$ and the Hubble rate at the end of inflation $H_*$ (which determines the inflaton reheating temperature). Moreover, the effective decay rate also depends on the parameters of the potential so that\footnote{
Although the Coleman-Weinberg correction \eq{deltaV} introduces a dependence on $m_h$, this turns out to be negligible because \eq{deltaV} turns out not to play a big role in the curvaton's evolution.
}
$\Gamma_{eff} = \Gamma_{eff}(\sigma_*,H_*,m_\sigma, g)$; the actual number can be computed by solving the equations of motion. The final curvature perturbation is then $\zeta=\zeta(\sigma_*,H_*,m_\sigma,\Gamma_{eff})$ with
\be{zeta}
\zeta=\zeta_G+\frac 35 f_{NL}\zeta_G^2+...\equiv \delta N =\frac{\partial N}{\partial \sigma}+\frac 12\frac{\partial^2 N}{\partial \sigma^2}+...
\ee
where the amplitude of $\zeta$ is fixed by $\zeta = 4.7 \times  10^{-5}$ \cite{Ade:2013zuv}. This constraint results in a 3D surface in the space of the parameters on which the non-gaussianity parameter $\fnl$ can be determined by numerically solving the equations of motion. In principle the curvaton contribution to other variables such as the spectral index of the primordial perturbations could be used to constrain the inflaton contibutions.

\subsection{Non-perturbative and perturvative decays}

Given the potential (\ref{V_0}), the curvaton can only decay into higgses, and this can happen via two different mechanisms. The first is is by a non-perturbative resonant process whereby the oscillating curvaton field excites higgs quanta. The second is by perturbative scatterings with the thermal background of higgses formed in inflaton decay. This has been discussed for the curvaton scenario in \cite{Enqvist:2012tc,Enqvist:2013qba}; for generic studies of the decay of a scalar field oscillating in  thermal background, see e.g. \cite{Drewes:2013iaa,decayofosc}. In addition, there exists a third possibility, decay through higher order non-renormalizable operators. The simplest possibility is to postulate decay by  gravitational strength interactions, which to lowest order are mediated via dimension-5 operators.  (Note that such operators  are likely to break global symmetries like  $\sigma\to -\sigma$.)
The total effective decay rate is thus given by
\be{totalrate}
\Gamma_{eff}=\Gamma_{NP} + \Gamma_5 + \Gamma_{pert} ~,
\ee
where $\Gamma_{NP}$ represents the contribution from non-perturbative decay, $\Gamma_5$ the contribution from dimension-5 operators, and $\Gamma_{pert}$ the contribution from perturbative scattering with the thermal bath. These are all discussed in the following sections.

\subsubsection{Non-perturbative decay}
Non-perturbative decay can occur and was discussed in detail for this model in \cite{Enqvist:2012tc,Enqvist:2013qba} (for earlier discussion of non-perturbative curvaton decay, see \cite{Enqvist:2008be,Chambers:2009ki,Kohri:2009ac,Sainio:2012rp}). For this, it is essential to take into account corrections due to the thermal background of the inflaton decay products. These affect both the curvaton potential and the dispersion relation of the higgs, which determines the nature and existence of the resonance \cite{Enqvist:2012tc}. The resonance occurs for certain momentum modes and can be labeled either as a broad or a narrow resonance. The resonance parameter is defined by
\be{q}
q(t) = \fb{g \Sigma(t)}{2m_\sigma}^2,
\ee
where $\Sigma(t)$ is the amplitude of the curvaton condensate at time $t$. A broad resonance has $q \gg 1$ whereas a narrow resonance has $q \ll 1$. The nature and effectiveness of the resonance depends on the curvaton initial conditions, its mass, its coupling to the higgs, and the subsequent evolution. In the absence of a thermal background, the curvaton would undergo broad resonance, and the timescale of energy transfer from the curvaton to the higgs would be fast. However, thermal corrections to the higgs mass can often block the curvaton decay for a long time \cite{Enqvist:2012tc}. As a consequence, the resonance parameter evolves, and typically the resonance is narrow at the time when the resonant production of higgses becomes possible.

The onset of narrow resonance occurs when the temperature satisfies
\begin{equation}\label{T_NR}
T_{\rm NR} \simeq \frac{m_\sigma}{g_T},
\end{equation}
where $g_T^2 \simeq 0.1$ \cite{Anderson:1991zb} is the effective thermal higgs coupling summed over all SM degrees of freedom. When the resonance is first unblocked, the center of the resonance band is near $k=0$ ($k$ is comoving momenta). As the thermal blocking becomes less important, the center of the band moves towards $(k/a) = m_\sigma$. The width of the resonance band is always given by $2q(t)$; this decreases with time. Thus, modes $k$ that are initially inside the resonance band will eventually leave it. Modes which are initially outside the band can enter and then leave. Because the narrow resonance is a resonant effect, the time a mode spends in the resonance band is important. Most energy is transferred to the higgs in modes which are initially outside the resonance band. The condition for efficient transfer of energy from the curvaton to the higgs is given by \cite{Enqvist:2013qba}
\begin{equation}\label{NRcond}
q^2(t_{NR}) m_\sigma t_{NR} \gtrsim 1,
\end{equation}
where $t_{NR}$ is the time when \eq{T_NR} is satisfied. If \eq{NRcond} is satisfied, then the resonance is efficient.
Although it is possible to determine the time when the resonance becomes unblocked and the curvaton condensate starts decaying into higgses, a rigorous numerical simulation is required to determine precisely when the decay stops, or the time when backreaction becomes important. This is beyond the scope of this project. However, it is known from simulations \cite{Kofman:1997yn, Podolsky:2005bw} that some fraction $F_{NP}$ of the condensate (or some non-relativistic particles) remains, typically between 1 and 10 \%. In a study of curvaton preheating it was found that about 5 \% of the condensate remains \cite{Chambers:2009ki}, and therefore for definiteness in this paper we choose $F_{NP}=0.05$ when \eq{NRcond} is satisfied and $F_{NP}=1$ when \eq{NRcond} is not satisfied. These assumptions have a negligible impact on the final results, because the curvature perturbation is dominated by the final decay. We have checked that our results are not sensitive to changes in $F_{NP}$.

To evaluate $\Gamma_{\rm NR}$, we need to numerically follow the evolution of both the curvaton field and the background, because $q(t)$ depends on the amplitude of the curvaton condensate. However, we can check the numerical results by making an analytic evaluation of $q(t_{\rm NR})$ in three separate cases. Applying the scaling law for energy densities of the curvaton and radiation, we find
\begin{equation}
\label{eq:q_case1}
q(t_{\rm NR}) \sim \fb{g}{g_T}^{3} \left( \frac{\sigma_\ast}{\Mpl} \right)^2 \left( \frac{\Mpl}{\Hinf} \right),
\end{equation}
for the case where oscillations begin immediately after inflation in the thermal potential \eq{thermalcorr}. In the second case there is a period of slow roll before oscillations begin in the thermal potential \eq{thermalcorr}, and we find
\begin{equation}
\label{eq:q_case2}
q(t_{\rm NR}) \sim \frac{g}{g_T^{3}} \left( \frac{\sigma_\ast}{\Mpl} \right)^2. 
\end{equation}
Finally, in the case where there is slow roll followed by oscillations in the quadratic potential \eq{V_0}, we find
\begin{equation}
\label{eq:q_case3}
q(t_{\rm NR}) \sim \frac{g^{2}}{g_T^{3}} \left( \frac{\sigma_\ast}{\Mpl} \right)^2 \left( \frac{\Mpl}{m_\sigma} \right)^{1/2}.
\end{equation}
The numerical and analytical results are in good agreement.

\subsubsection{Gravitational strength decay}

In addition to the renormalizable coupling to the higgs, higher order effective operators are expected to exist since there is no symmetry that would explicitly forbid them. These include the curvaton coupling to standard model fermions $f$ through  dimension-5 operators such as
\be{dim5}
{\mathcal L}_5\propto \frac{1}{M_P}\sigma \bar f \Phi f~.
\ee
This is just one example; there would also be $d=5$ operators involving gauge fields. On dimensional grounds and neglecting possible coefficients, the $d=5$ operators in combination yield a curvaton decay rate of
\be{decrate5}
\Gamma_5\approx \frac{m_\sigma^3}{M_P^2}~.
\ee
Note that we have chosen $M_P$ as the scale of the higher order operators, and therefore no new parameters have been introduced. In principle, the scale is unknown{\footnote{
However, it should be much larger than the curvaton mass, which should be smaller than the Hubble rate during inflation, which could be as high as $10^{12}$ GeV.
}. Since we implicitly assume that, apart from the curvaton, there is no new physics beyond the standard model and Einstein gravity, choosing $M_P$ as the scale of non-renormalizable physics seems justified.

Although the decay rate (\ref{decrate5}) is Planck suppressed, it can nevertheless be more important than the non-perturbative decay. Moreover, after backreaction shuts down the resonant production of higgses, in the minimal scenario presented here it is the higher-order operators that will be responsible for completing the curvaton decay.

If dimension-5 operators were forbidden for some reason, the effective decay rate by dimension-6 operators would be (again on dimensional grounds and neglecting coefficients)
\be{decrate6}
\Gamma_6\approx \frac{m_\sigma^5}{M_P^4}~.
\ee

We require the curvaton to decay before BBN to avoid spoiling the predictions \cite{Lyth:2003dt}. Specifically, it should decay before the neutrinos decouple at $T = 4\MeV$ \cite{Enqvist:1991gx}. These limits are found by assuming $H(t) \propto T^2$ and calculating when $H(t) = \Gamma_5$ or $\Gamma_6$. For the dimension-5 operator, this corresponds to requiring $m_\sigma \gtrsim 8\times 10^4\GeV$.  However, for the dimension-6 operator, the condition on $m_\sigma$ is much stronger, giving $m_\sigma \gtrsim 4\times 10^{10}\GeV$. This would only allow a small window of curvaton masses; the upper bound is given by $m_\sigma < H_*$. However, we presume that the dimension-5 operators are not forbidden, and thus a substantial amount of parameter space remains.

\subsubsection{Perturbative scatterings with the thermal background}
The presence of a thermal bath means that decay processes for the curvaton include perturbative scatterings with the thermal background, the importance of which were first pointed out in \cite{Dimopoulos:2003ss}. We calculate the decay width for this, also including $1\to 3$ particle decays and production of $k=0$ modes {\em from} the thermal background. One finds \cite{Elmfors:1993re}
\be{new1}
\Gamma_{pert} = \frac{1}{576\pi} \frac{g^4 T^2}{m_\sigma(T)}.
\ee
We now consider the two cases for initial effective mass at the temperature $T_*$, where $T_*$ is the reheating temperature assuming instant reheating. If $m_\sigma(T_*) = m_\sigma$, then because $H(t)\propto T^2 / \Mpl$, decay by these perturbative interactions either never occurs, or occurs immediately after the thermal background has been produced. If curvaton decay occurs immediately, then the curvaton model is ruled out because the relative fraction of the curvaton energy density has not had time to increase to satisfy the normalization $\zeta = 4.7 \times 10^{-5}$. This rules out large $g$, and depends on $m_\sigma$. If instead $m_\sigma(T_*) = gT/\sqrt{12}$, then we should determine whether these perturbative interactions can occur before $T \simeq m_\sigma / g$, which marks the transition to $m(T) = m_\sigma$. In that case, the curvaton decays while both $\rho_\sigma$ and $\rho_{rad}$ scale $\propto a^{-4}$. This means that the relative fraction of the curvaton energy density cannot grow, and that the curvaton model is ruled out in this region. The condition for this to occur is given by
\be{new2}
g \geq 4.9 g_*^{1/8} \fb{m_\sigma}{\Mpl}^{1/4},
\ee
where $g_*=107.75$ is the effective number of degrees of freedom.

\section{The curvature perturbation and non-Gaussianity}
\label{sec:perturbation}

Now we discuss the curvature perturbation in the model.
We adopt the $\delta N$ formalism \cite{starob85,ss1,Sasaki:1998ug,lms} to calculate the curvature perturbation
up to the 3rd order:
\begin{equation}
\zeta
=  N_\sigma \delta \sigma_\ast
+
\frac12 N_{\sigma\sigma} \left( \delta \sigma_\ast \right)^2
+
\frac16 N_{\sigma\sigma\sigma} \left( \delta \sigma_\ast \right)^3,
\end{equation}
where $N_\sigma =d N / d \sigma_\ast$ and so on.  Then the power, bi- and tri-spectra are defined respectively as
\begin{eqnarray}
\left\langle \zeta (\bm{k}_1)  \zeta (\bm{k}_2) \right\rangle
& = &
(2\pi)^2 P_\zeta ( k_1) \delta (\bm{k}_1 + \bm{k}_2 ), \\
\left\langle \zeta (\bm{k}_1)  \zeta (\bm{k}_2) \zeta (\bm{k}_3) \right\rangle
&=&
(2\pi)^2 B_\zeta ( k_1, k_2, k_3) \delta (\bm{k}_1 + \bm{k}_2  +  \bm{k}_3), \\
\left\langle \zeta (\bm{k}_1)  \zeta (\bm{k}_2) \zeta (\bm{k}_3) \zeta (\bm{k}_4) \right\rangle
& = &
(2\pi)^2 T_\zeta ( k_1, k_2, k_3, k_4) \delta (\bm{k}_1 + \bm{k}_2 + \bm{k}_3 + \bm{k}_4 ), \\
\end{eqnarray}
where $B_\zeta$ and $T_\zeta$ are given by
\begin{eqnarray}
B_\zeta (k_1,k_2,k_3)
&=&
\frac{6}{5} f_{\rm NL}
\left(
P_\zeta (k_1) P_\zeta (k_2)
+ P_\zeta (k_2) P_\zeta (k_3)
+ P_\zeta (k_3) P_\zeta (k_1)
\right), \\
\label{eq:fNL}
T_\zeta (k_1,k_2,k_3,k_4)
&=&
\tau_{\rm NL} \left(
P_\zeta(k_{13}) P_\zeta (k_3) P_\zeta (k_4)+11~{\rm perms.}
\right) \nonumber \\
&&
+ \frac{54}{25} g_{\rm NL} \left( P_\zeta (k_2) P_\zeta (k_3) P_\zeta (k_4)
+3~{\rm perms.} \right),
\label{eq:tau_gNL}
\end{eqnarray}
where $k_{13} = | \vec{k}_1 + \vec{k}_3 |$. When the curvature perturbations are generated from the curvaton field alone,
the power spectrum is given by
\begin{equation}
P_\zeta (k) = N_\sigma^2 P_{\delta \sigma} (k),
\end{equation}
where $P_{\delta \sigma}$ is the power spectrum for fluctuations of the curvaton field $\delta \sigma$, which
is given by
\begin{equation}
P_{\delta \sigma} (k) = \frac{2 \pi^2}{k^3} \mathcal{P}_{\delta \sigma} (k).
\end{equation}
Here $\mathcal{P}_{\delta \sigma} = (H/ 2\pi)^2$ with $H$ the Hubble parameter at the time
of horizon crossing for a given mode $k$.

The non-linearity parameters $f_{\rm NL}$ and $g_{\rm NL}$ are given by
\begin{eqnarray}
\frac56 f_{\rm NL}  &=&
\frac{N_{\sigma\sigma}}{N_\sigma^2},\\
\frac{54}{25} g_{\rm NL}  &=&
\frac{N_{\sigma\sigma\sigma}}{N_\sigma^3}
\end{eqnarray}
whereas  $\tau_{\rm NL}= \left( 5 f_{\rm NL}/6 \right)^2$.

For the curvaton, $\zeta$ can be expressed as \cite{Sasaki:2006kq}
\begin{eqnarray}
\label{eq:zeta_cur}
\zeta_{\rm cur} &=&
\frac{2}{3} r_{\rm dec} \frac{\sigma'_{\rm osc}}{\sigma_{\rm osc}}  \delta \sigma_\ast
+
\frac{1}{9} \left[ 3r_{\rm dec}\left(
1 +
\frac{\sigma_{\rm osc} \sigma_{\rm osc}^{\prime\prime}}{\sigma_{\rm osc}^{\prime 2}}
\right)
- 4 r^2_{\rm dec} -2  r^3_{\rm dec}
\right]
\left( \frac{\sigma'_{\rm osc}}{\sigma_{\rm osc}} \right)^2  (\delta \sigma_\ast )^2 \notag \\
&&+
\frac{4}{81} \left[
\frac{9r_{\rm dec}}{4}  \left(
\frac{\sigma_{\rm osc}^2 \sigma_{\rm osc}^{\prime\prime\prime}}
{\sigma_{\rm osc}^{\prime 3}}
+
3\frac{\sigma_{\rm osc} \sigma_{\rm osc}^{\prime\prime}}{\sigma_{\rm osc}^{\prime 2}}
\right)
-9r^2_{\rm dec}
\left(
1
+
\frac{\sigma_{\rm osc} \sigma_{\rm osc}^{\prime\prime}}{\sigma_{\rm osc}^{\prime 2}}
\right)
\right. \notag \\
&&
\left.
+\frac{r^3_{\rm dec}}{2} \left(
1
-
9\frac{\sigma_{\rm osc} \sigma_{\rm osc}^{\prime\prime}}{\sigma_{\rm osc}^{\prime 2}}
\right)
+10r^4_{\rm dec} + 3r^5_{\rm dec}
\right]
\left( \frac{\sigma'_{\rm osc}}{\sigma_{\rm osc}} \right)^3  (\delta \sigma_\ast )^3~,
\end{eqnarray}
where $\sigma'_{\rm osc} = d \sigma_{\rm osc} / d \sigma_\ast$, and $\sigma_{\rm ocs}$ indicates the field value of $\sigma$ at the time when the curvaton begins its oscillation under the bare mass term in the potential \eqref{V_eff}. The parameter
$r(t)$ is defined by
\begin{equation}
\label{eq:def_r}
r (t)
\equiv \frac{3 \rho_\sigma (t) }{4 \rho_{\rm rad} (t) + 3\rho_\sigma (t) },
\end{equation}
which roughly corresponds to the ratio of the curvaton energy density to the total.
In  particular, we define $r_{\rm dec} \equiv r (t=t_{\rm dec})$, where $t_{\rm dec}$ is the time of curvaton decay.
Notice that we assume that the narrow resonance  converts the energy of the curvaton into
radiation only partially. Thus, the curvaton totally decays into radiation  only after the perturbative decay.

\begin{figure}[htbp]
  \begin{center}
     \includegraphics{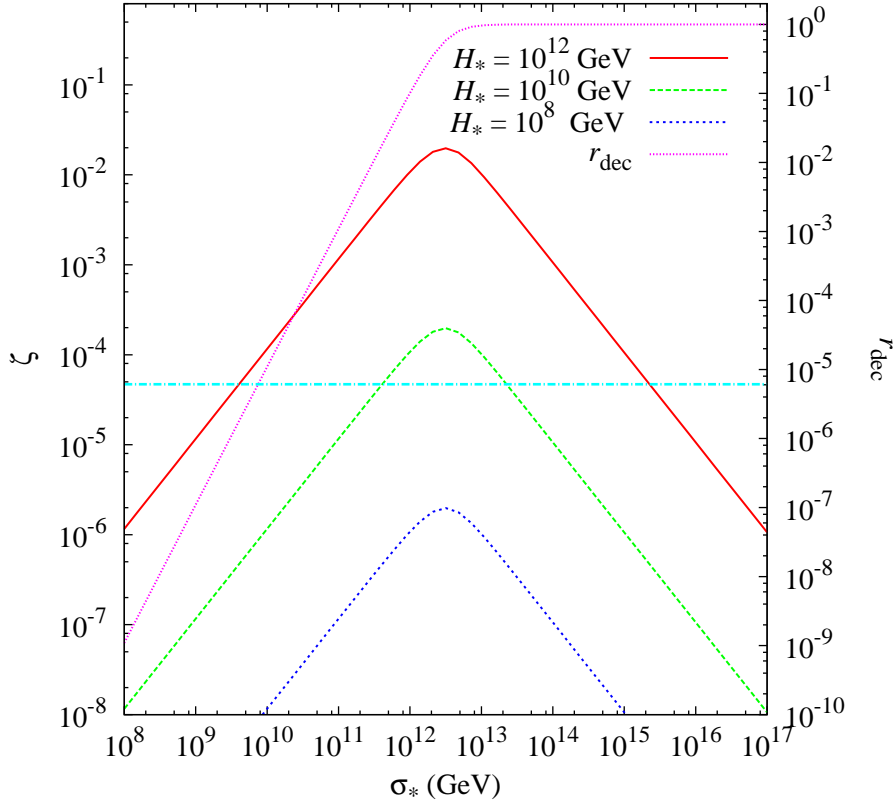}
  \end{center}
  \caption{The dependence of $\zeta$ on the inflation scale  $\Hinf$. Here we assume that
  $m_\sigma = 10^{6}$~GeV and $g=10^{-13}$. The horizontal line marks the observed  $\zeta = 4.7 \times  10^{-5}$. Also shown is the dependence of  $r_{\rm dec}$ on the initial curvaton value  $\siga$.}
  \label{fig:zeta_H}
\end{figure}
\begin{figure}[htbp]
  \begin{center}
     \includegraphics{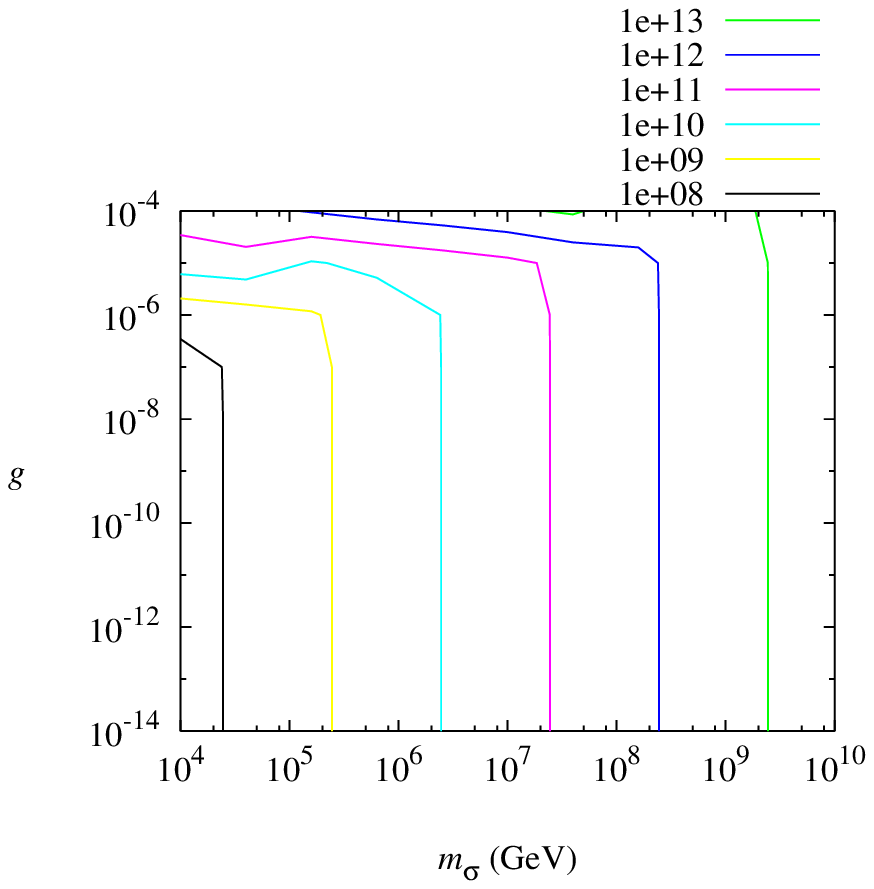}
  \end{center}
  \caption{Contours of $\siga$ in the case of Solution 2 that give the correct amplitude $\zeta = 4.7 \times 10^{-5}$, with small $m_\sigma$ corresponding to small $\sigma_*$. Here we assume that   $\Hinf=10^{12}$~GeV.}
  \label{fig:sig_m_g_plane}
\end{figure}

For the analysis presented in this paper, we follow the the number of $e$-folds numerically from just after inflation.
We assume that the inflaton decays into radiation just after inflation ends, and thus we follow the energy density of
radiation and the curvaton field from the end of inflation. The number of $e$-folds $N$ can be obtained by solving the following set of equations:
\begin{eqnarray}
&& \frac{d \rho_r}{dt}  + 4 H \rho_r  =  \Gamma \rho_\sigma, \\
&& \frac{d^2 \sigma}{dt^2} + (3 H + \Gamma)  \frac{d \sigma}{d t}  + \frac{dV}{d\sigma} =   0, \\
&& H^2 = \left( \frac{1}{a} \frac{da}{dt} \right)^2 = \frac{1}{3 M_P^2} (\rho_r + \rho_\sigma),
\end{eqnarray}
where the curvaton decay rate $\Gamma$ is given by Eq.~\eqref{totalrate}. Note that $\Gamma_{\rm NP}$ is zero before the narrow resonance is unblocked, which occurs at $t = t_{\rm NR}$ (see \eq{T_NR}). The initial velocity of $\sigma$ is given by the slow-roll solution.

We follow the above set of equations numerically until the time when $H = \Gamma_5$ is satisfied,  when
we can evaluate $N$ and its derivatives. Then we can  calculate the power spectrum and the 
non-linearity parameters, which are presented in the following section.

To obtain  $\zeta = 4.7 \times  10^{-5}$  requires specific values of $\siga$. There are two possible values of $\siga$ which satisfy this requirement, as shown in \fig{fig:zeta_H}. We denote them as Solution 1 and Solution 2. Solution 1 always has $r_{\rm dec}\simeq 1$ and corresponds to what in the literature is called the dominant curvaton. Because the non-Gaussianity parameter $\fnl\propto 1/r_{\rm dec}$, there is therefore no constraint from the recent Planck data (the two-sigma upper limit is $\fnl \leq 14.3$).

Solution 2, which in   Fig.~\ref{fig:zeta_H} corresponds to the smaller value of $\siga$, often leads to a subdominant curvaton, but for a range in $\Hinf$ can also give rise to a slightly dominant curvaton. This can be seen in Fig.~\ref{fig:zeta_H}, where $r_{\rm dec}$ associated with Solution 2  is seen to approach 1 as $\Hinf$ decreases. In this case the Planck limit on $\fnl$ can be expected to constrain the parameter space. Note however that the parameter space of Solution 2 resides inside the parameter space of Solution 1. For illustration, in Fig.~\ref{fig:sig_m_g_plane} we show the contours of $\siga$ for Solution 2 that lead to the observed curvature perturbation. The figure is for fixed $\Hinf$; the contours look different for different Hubble rates. For Solution 1,
$\siga \simeq 10^{15}$ GeV. This is slightly modified when the thermal correction initially dominates the potential, for large $g$ and small $m_\sigma$. The Colemann-Weinberg potential never dominates the potential, and we find that it also has a negligible affect on the non-Guassianity parameters.
We have also checked that the Coleman-Weinberg potential does not make the curvaton heavy during inflation unless $g^2 \gtrsim 24\pi^2\zeta$ or $g\gtrsim 0.5$, which is outside the region of interest.

\section{Results}
\label{sec:results}
We present results first for Solution 1, where the curvaton is always dominant at the final decay, and discuss separately Solution 2. We fix the value of $\sigma_*$ such that the observed value of $\zeta$ is produced. Various constraints apply to the parameter space, including requiring decay before BBN.

We find that if the condition \eq{new2} is satisfied, then the curvaton decays immediately via interactions with the thermal bath and the model does not work. However, these interactions with the thermal bath are not important in the rest of the parameter space. If \eq{new2} is not satisfied, then for large $m_\sigma$, the dimension-5 process occurs before the narrow resonance is unblocked. If $m_\sigma$ is smaller, then the narrow resonance can occur first. This could either be efficient (transferring 95\% of the energy out of the curvaton) or inefficient. However, in both cases the final curvaton decay is due to the dimension-5 operator. If the curvaton is dominant at its decay, then the predictions are not particularly affected by whether there was first a narrow resonance. If the curvaton is either subdominant or only slightly dominant at decay, then the narrow resonance can affect the non-Gaussianity.

\subsection{Solution 1}
\begin{figure}[htbp]
  \begin{center}
     \includegraphics{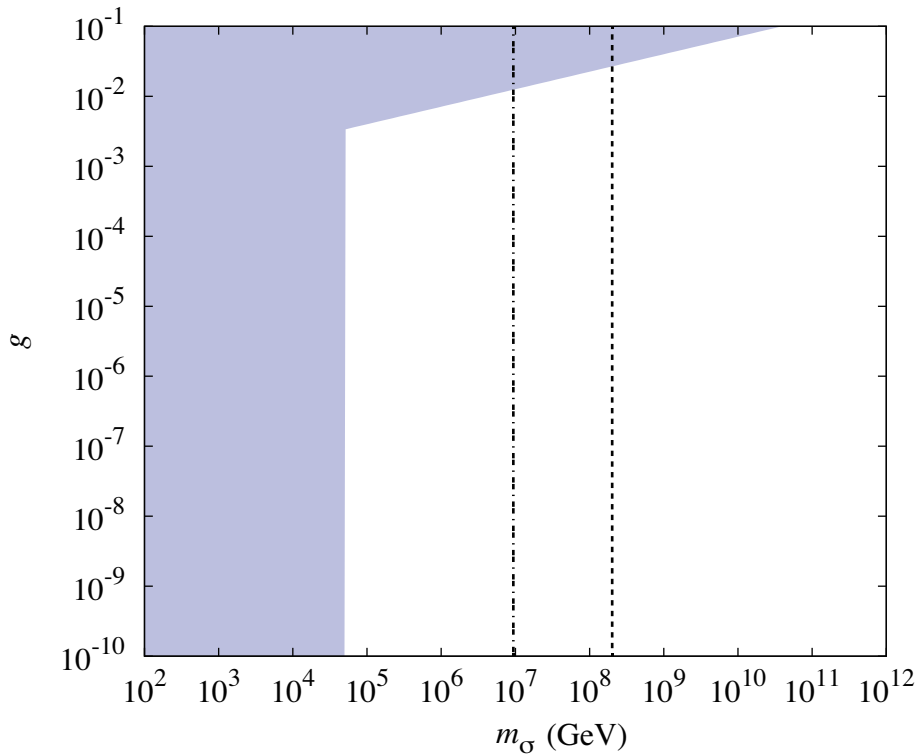}
  \end{center}
  \caption{Parameter space for Solution 1, the dominant curvaton (white is allowed). Dashed lines show how the lower limits on $m_\sigma$ would increase if WIMP dark matter froze out at $10\GeV$ (lower) or $1\TeV$ (upper). Note that the upper bound $m_\sigma < H_*$ is not shown because it depends on $H_*$; also not shown for this reason is the effect of the Coleman-Weinberg potential.}
  \label{fig:dom}
\end{figure}

For Solution 1 the curvaton is always dominant at decay and non-Gaussianity is small. The constraints on the model are shown in \fig{fig:dom}, where the allowed area is white. There is also an upper limit on $m_\sigma$ that comes from requiring the curvaton to be light during inflation, i.e.\ $m_\sigma < H_*$. The upper limit on $g$ comes from requiring the interactions with the thermal bath not to immediately destroy the curvaton condensate. The lower limit on $m_\sigma$ comes because the final decay width is determined solely by the curvaton mass. To avoid spoiling BBN the curvaton must decay sufficiently early, before $T \simeq 4 \MeV$ (this is when $\nu_\mu$ and $\nu_\tau$ decouple \cite{Enqvist:1991gx}). There is thus a lower limit on the curvaton mass. If dark matter is a thermal relic, then large isocurvature would be predicted if the curvaton decays after the dark matter freezes out. This is in contrast with observations. Thus, the dot-dashed line in \fig{fig:dom} shows the lower limit on $m_\sigma$ for a WIMP dark matter model that freezes out at $10\GeV$ while the dashed line is for a WIMP with freeze-out at 1 TeV. However, other dark matter models do not impose such a bound. A determination of the properties of dark matter will allow this constraint to be properly calculated. There could also be a lower bound from requiring the curvaton not to spoil baryogenesis. However, this is also model dependent.

\begin{figure}[htbp]
  \begin{center}
     \includegraphics[width=0.77\textwidth]{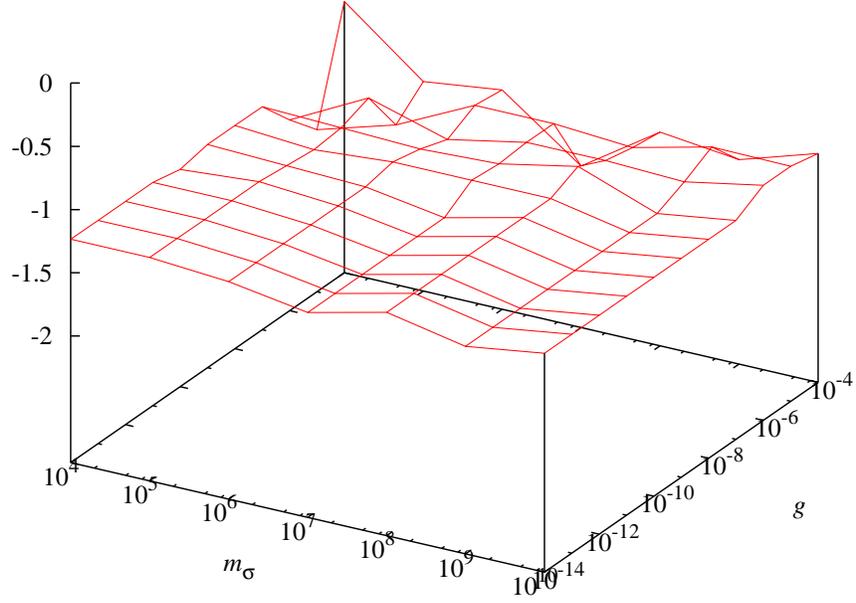}
  \end{center}
  \caption{$\fnl$ for Solution 1, shown for $H_* = 10^{12}\GeV$.}
  \label{fig:dom_NG}
\end{figure}

The Coleman-Weinberg potential is initially dominant for Solution 1 if $g\gtrsim 0.1 \sqrt{m_\sigma/H_*}$. \fig{fig:dom} is plotted analytically and does not include this effect of the Coleman-Weinberg potential, because it depends on $H_*$. However, we expect that including it only changes the initial field value necessary to match observations, $\sigma_*$. The remaining figures show numerical results including all terms in the potential. 

Although neither the Coleman-Weinberg potential nor the thermal correction dominate in most of the parameter space, they could still introduce non-Gaussianity. We have checked this carefully and found that the variations are small but possibly interesting. The precise predictions for $\fnl$  for Solution 1 are shown in \fig{fig:dom_NG} for the case of $H_* = 10^{12}\GeV$. We find a small, negative $\fnl$ with values in the range $-1.5 \lesssim \fnl \lesssim 0$. We have also checked that for Solution 1,  $\gnl$  is unobservable with $\gnl\lesssim 100$. We note that in all cases, both the curvature perturbation and the non-Gaussianity are produced at the second (final) decay of the curvaton. Although large non-Gaussianity can occur after the first decay, this is then diluted by the subsequent expansion. In all cases where the curvaton is dominant at decay, the first decay only has a negligible impact on the parameters. This is under the assumption that the resonance can never destroy all of the condensate. This has not been explicitly shown with a lattice simulation for our model, but is based on similar models in the literature.

\begin{figure}[htbp]
  \begin{center}
     \includegraphics[width=0.4\textwidth]{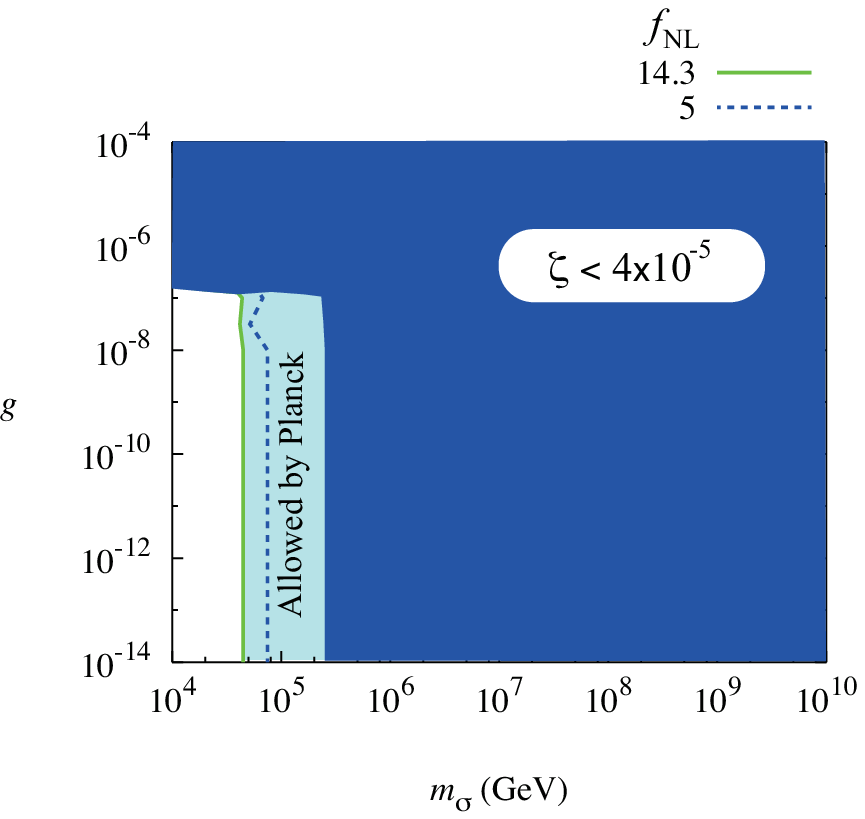}
    \includegraphics[width=0.4\textwidth]{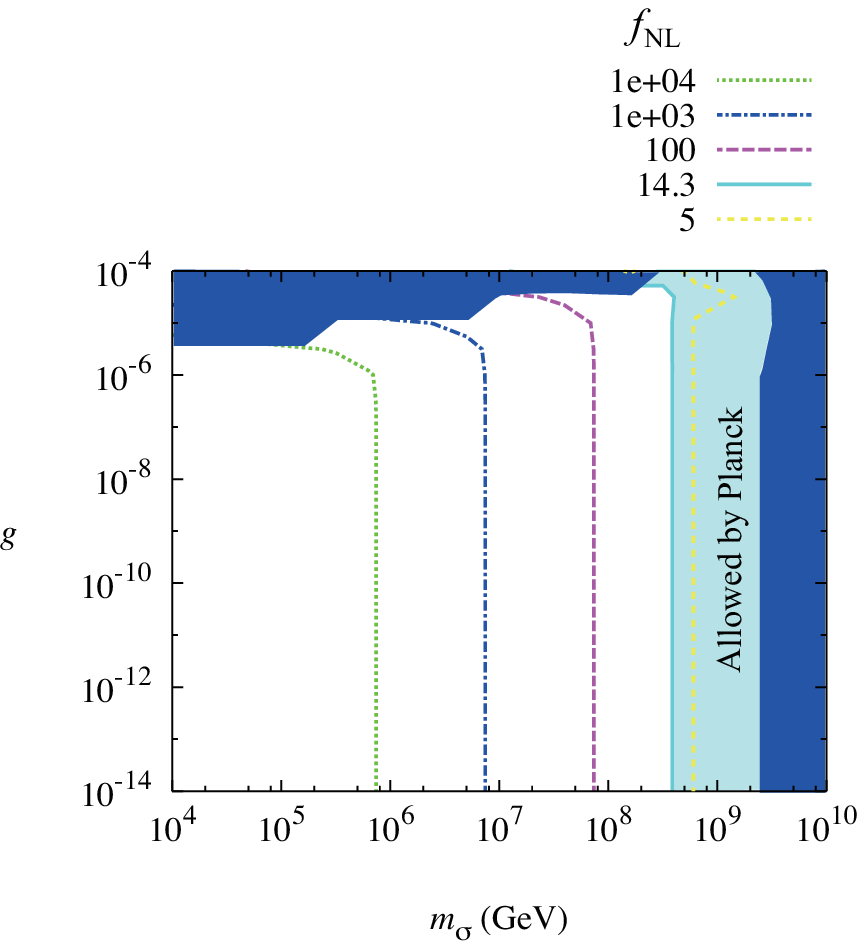}
  \end{center}
  \caption{The allowed parameter space (pale blue / light grey) for Solution 2, with contours of $\fnl$ shown. Left: $H_* = 10^{9}\GeV$; right: $H_* = 10^{11}\GeV$.}
  \label{fig:subdom_NG}
\end{figure}

\subsection{Solution 2}

When the curvaton is subdominant at both decays, large non-Gaussianity can be produced and the parameter space can be restricted by the Planck limit $\fnl \leq 14.3$. We then find that Solution 2 exists in the range $10^9~{\rm GeV}\lesssim H_*\lesssim 10^{11}~{\rm GeV}$. However, most of the parameter space of Solution 2 has $\fnl$ that is not observable, e.g.\ $\fnl \lesssim 5$. This is evident from
\fig{fig:subdom_NG}, which shows contours of $\fnl$ for Solution 2 with $\sigma_*$ fixed to give the correct amplitude of $\zeta$, together with the bounds described above on $g$ and $m_\sigma$. The allowed space of parameters for Solution 2 is contained within the parameter space of Solution 1 depicted in \fig{fig:dom}. If $\fnl$ were to be observed, it would imply a curvaton mass in the range $8\times 10^4\GeV \lesssim m_\sigma \lesssim 5\times 10^9\GeV$, as can be seen from \fig{fig:subdom_NG}.

\section{Conclusion}

We have presented the minimal curvaton-higgs (MCH) model and attempted to include all interactions of the model, including dimension-5 gravitationally suppressed operators. At tree-level, the curvaton is stable. However the decay mechanisms that do exist are (i) interactions with the thermal background that exists because of the earlier inflaton decay, (ii) non-perturbative decay via narrow resonance, which is never completely efficient, and (iii) decay via the dimension-5 operator. Although in principle all three decay mechanisms could contribute to the observed curvature perturbation, we find that in the allowed parameter space only the dimension-5 operators impact the predictions of the model.

We  demonstrated that the MCH model is viable and consistent with various constraints, and that typically the predicted non-Gaussianities are small. However, the possibility of observable $\fnl$ still exists. More specifically, we found that the implications of the MCH model are: (a) the curvaton mass is greater than $8\times 10^4\GeV$, (b) the curvaton-higgs coupling is constrained, depending on the mass, but $g<10^{-3}$ is always allowed, (c) if $\fnl$  were to be observed, then $m_\sigma \lesssim 5\times 10^9\GeV$, (d) if dark matter is observed and the freeze-out temperature is $10\GeV$ (1\TeV), then the curvaton mass should be greater than $10^7\GeV$ ($10^8\GeV$). If $H_*$ is determined to be low, and dark matter is found to be thermal relic, then the parameter space will become strongly constrained.

The MCH implementation of the curvaton scenario is important because it presents a robust model for the origin of the primordial perturbation where the particle content is fully specified. The dynamics of the model (such as the effective decay width) are not free parameters, but are given by the parameters in the Lagrangian. The model is testable with current and upcoming data and could also have implications for LHC physics. One interesting possibility would be a connection with the so-called Higgs-portal models of dark matter, where the standard model higgs is coupled to a real singlet scalar \cite{portal}.

\section*{Acknowledgements}
We thank Sami Nurmi and Dmitry Podolsky for comments on the fraction of the condensate remaining  in preheating. TT would like to thank the Helsinki Institute of Physics for the hospitality during a visit, where part of this work was completed. The work of TT is partially supported by the Grant-in-Aid for Scientific research from the Ministry of Education, Science, Sports, and Culture, Japan, No.~23740195. KE is supported by the Academy of Finland grant 1218322; RL was supported by the Academy of Finland grant 1263714.

\end{document}